\documentclass{article}
\usepackage{spconf,amsmath,epsfig}
\usepackage{subcaption}
\usepackage{float}
\usepackage{amsfonts}
\usepackage{xcolor}
\usepackage{multirow}
\usepackage{hyperref}

\let\OLDthebibliography\thebibliography
\renewcommand\thebibliography[1]{
  \OLDthebibliography{#1}
  \setlength{\parskip}{0pt}
  \setlength{\itemsep}{0pt plus 0.3ex}
}

\begin{document}\sloppy

% Example definitions.
% --------------------
\def\x{{\mathbf x}}
\def\L{{\cal L}}

\title{Hierarchical and Contrastive Representation Learning for Knowledge-aware Recommendation}

\name{Bingchao Wu\textsuperscript{1,5}, Yangyuxuan Kang\textsuperscript{2}, Daoguang Zan\textsuperscript{1,5}$^{\dagger}$, Bei Guan\textsuperscript{4,5}$^{\dagger}$\thanks{$\dagger$ Corresponding author.}, Yongji Wang\textsuperscript{3,4,5}}
\address{\textsuperscript{1} Collaborative Innovation Center, Institute of Software, Chinese Academy of Sciences, Beijing, China \\
        \textsuperscript{2} Intel Labs China, Beijing, China \\
	\textsuperscript{3} State Key Laboratory of Computer Science, Institute of Software, \\ Chinese Academy of Sciences,  Beijing, China \\
        \textsuperscript{4} Integrative Innovation Center, Institute of Software, Chinese Academy of Sciences, Beijing, China \\
        \textsuperscript{5} University of Chinese Academy of Sciences, Beijing, China\\
        \{paulpigwbc@outlook.com, yangyuxuan.kang@intel.com, \\ \{daoguang,guanbei\}@iscas.ac.cn, ywang@itechs.iscas.ac.cn \}
}

\maketitle

\begin{abstract}
Incorporating knowledge graph into recommendation is an effective way to alleviate data sparsity.
Most existing knowledge-aware methods usually perform recursive embedding propagation by enumerating graph neighbors. 
However, the number of nodes’ neighbors grows exponentially as the hop number increases, forcing the nodes to be aware of vast neighbors under this recursive propagation for distilling the high-order semantic relatedness.
This may induce more harmful noise than useful information into recommendation, leading the learned node representations to be indistinguishable from each other, that is, the well-known over-smoothing issue. 
To relieve this issue, we propose a Hierarchical and CONtrastive representation learning framework for knowledge-aware recommendation named HiCON.
Specifically, for avoiding the exponential expansion of neighbors, we propose a hierarchical message aggregation mechanism to interact separately with low-order neighbors and meta-path-constrained high-order neighbors. 
Moreover, we also perform cross-order contrastive learning to enforce the representations to be more discriminative.
Extensive experiments on three datasets show the remarkable superiority of HiCON over state-of-the-art approaches. 
\end{abstract}

\begin{keywords}
Knowledge-aware recommendation, hierarchical message aggregation, contrastive learning
\end{keywords}

\section{Introduction}
\label{sec:intro}
Data sparsity is a significant challenge in the recommendation system.
A natural way to alleviate this challenge is incorporating external side information~\cite{side_information}. 
Knowledge graph (KG), a common and essential source including real-world facts, can assist the recommendation system in learning better representations via its rich structural and semantic information~\cite{TransE}. 
Consequently, it is widely used in existing methods to alleviate the data sparsity issue~\cite{KGAT,KGIC}. 
\begin{figure}[t]
  \centering
  \begin{subfigure}{0.24\textwidth}
    \hspace{-0.0in}\includegraphics[scale=0.24]{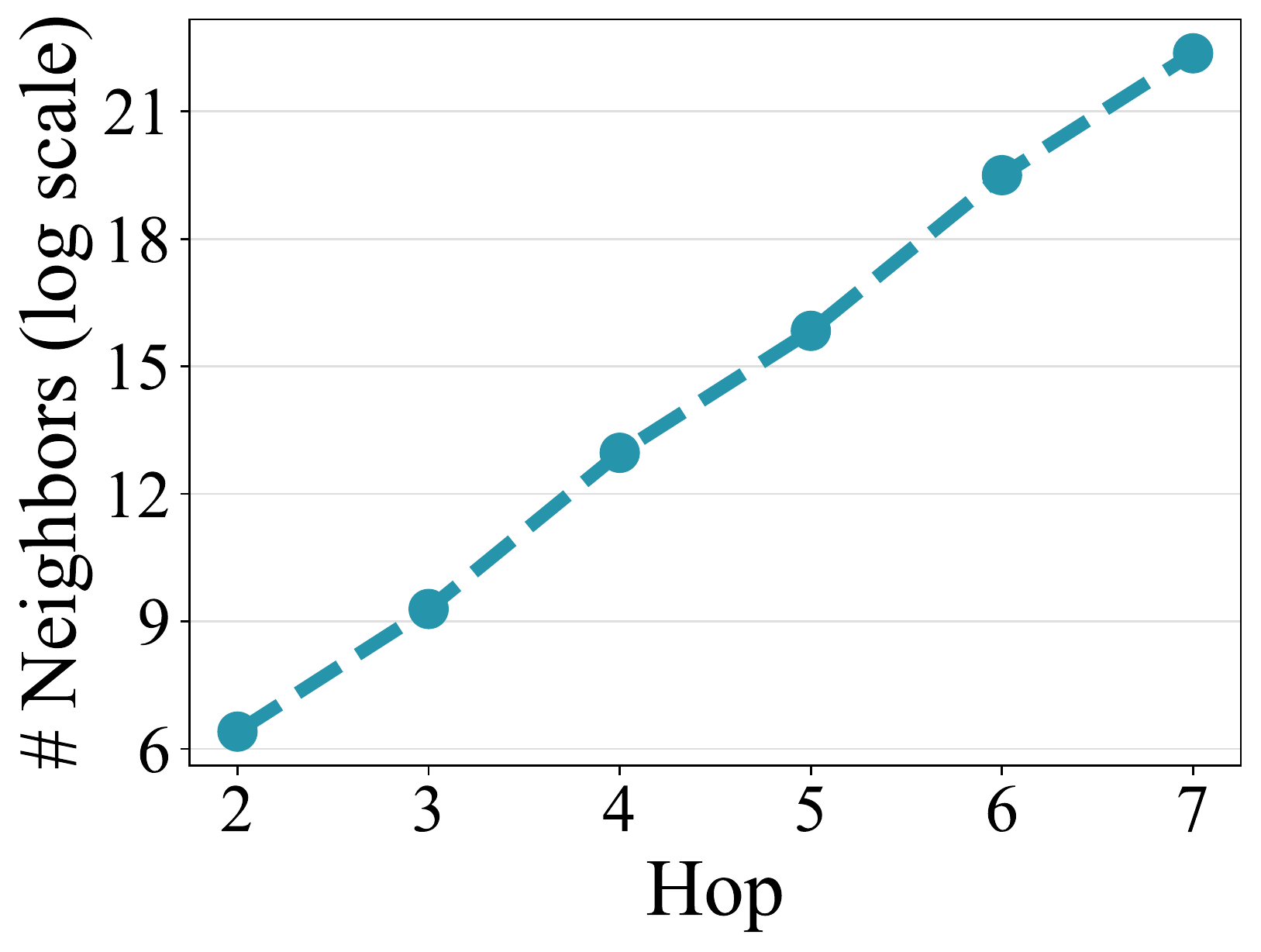}
    \caption{Average number of interacted neighbors.}
    \label{Fig:neighbor_number}
  \end{subfigure}
  \begin{subfigure}{0.23\textwidth}
    \hspace{-0.16in}
    \includegraphics[scale=0.24]{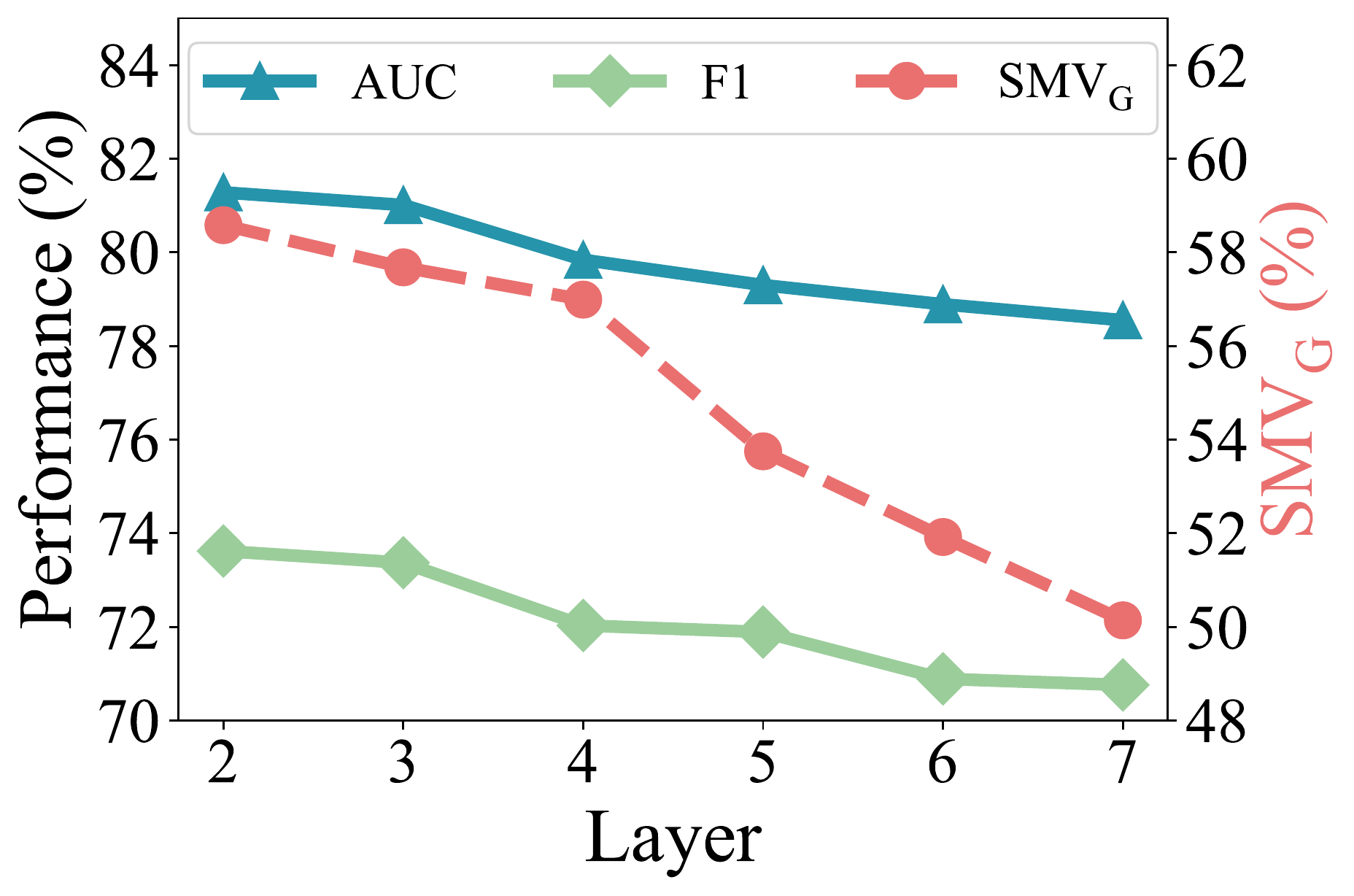}
    \caption{Performance and smoothness metric of KGAT.}
    \label{Fig:over_smoothing}
  \end{subfigure}
  \vspace{-0.05in}
  \caption{Illustration of the fact that graph neighbors grow exponentially and the over-smoothing issue. Figure (a) counts the average number of interacted neighbors per user at different hops. Since KGAT has to stack as many layers as hops for capturing more complicated semantic relatedness, figure (b) reports recommendation performance (i.e., AUC and F1) and smoothness (i.e., SMV$_G$) at the different number of layers of KGAT (AUC, F1, SMV$_G$: the higher, the better).}
  \vspace{-0.18in}
  \label{Fig:issues}
\end{figure}
Most earlier works~\cite{CKE,embedding_based_DKN} utilize knowledge graph embedding (KGE) methods, such as TransR~\cite{TransR} and TransD~\cite{TransD}, to pre-train entity embeddings based merely on a single KG triplet (consists of two entities and a connected relation) for enriching item-side semantic representations.
However, due to modeling the triplets independently, they may insufficiently capture the high-order relatedness among nodes~\cite{KGIC}.

Recently, most existing studies develop knowledge-aware recommendation methods based on graph neural networks (GNN) to explore high-order relatedness for accurate recommendation, such as KGAT~\cite{KGAT} and KGCN~\cite{KGCN}. 
A typical GNN paradigm usually performs recursive message propagation by enumerating graph neighbors over the unified graph consisting of the user-item bipartite graph and item-side knowledge graph.
There exists a fact that many nodes are aware of exponentially growing neighbors over the unified graph as the number of hops increases (shown as Figure~\ref{Fig:neighbor_number}). 
Thus, to capture the high-order semantic relatedness with these neighbors, this GNN paradigm has to force the nodes to be aware of the information of all exponentially expanding neighbors because of its recursive propagation mechanism.
This may bring in more infernal noise than useful information for node representation learning, which greatly harms the recommendation performance~\cite{Ripplenet}.
Several existing works attribute this phenomenon to the over-smoothing issue~\cite{oversmoothing_metric}. 
As shown in Figure~\ref{Fig:over_smoothing}, we take KGAT (a representative knowledge-aware recommendation model) as an example to intuitively describe this issue.
To reflect the overall smoothness of node representations, we employ SMV$_G$ proposed in 
DAGNN~\cite{oversmoothing_metric}, which is the average pairwise Euclidean distance of two-node representations. 
We find that the overall smoothness metric SMV$_G$ of node representations learned from KGAT continuously degrades 
along with the increase of the number of message propagation layers, leading to the degradation of performance on AUC and F1. 
  
In this paper, we propose an effective framework named HiCON for knowledge-aware recommendation. 
It alleviates the over-smoothing issue in two aspects: 
1) selecting and propagating a bundle of valuable neighbors to the central nodes rather than considering all exponentially increasing neighbors to reduce noise disturbance; 
2) enhancing the self-discrimination of node representations in the latent space. 
For the first aspect, we propose a hierarchical message aggregation mechanism consisting of two parts, i.e., low- and high-order message aggregations, to avoid interacting with exponentially expanding neighbors.
The low-order aggregation enriches the node representations at the low level by aggregating local graph neighbors over the unified graph.
The high-order aggregation aims to learn high-level node representations by first selecting valuable high-order neighbors guided by well-designed meta-paths and then propagating the information of selected neighbors to the central node. 
For the second aspect, to enhance the discriminativeness of learned node representations, we perform cross-order contrastive learning between low- and high-level semantic representations derived from the message aggregation process.
Extensive experiments on three benchmark datasets show that HiCON can greatly improve the recommendation performance and meanwhile alleviate over-smoothing.  

Overall, our work makes two major contributions: 
1) We propose a  hierarchical and contrastive representation learning framework for knowledge-aware recommendation, which alleviates over-smoothing to improve performance from two aspects: i.e., avoiding exponential expansion of interacted neighbors and enhancing the discriminativeness of node representations. 
2) Extensive experiments on three datasets validate the effectiveness of HiCON in providing accurate recommendations and the ability to alleviate over-smoothing.

\section{Related Work}
\noindent
\textbf{Knowledge-aware Recommendation.}
Incorporating the structural and semantic information of knowledge graph (KG) into recommendation is effective to alleviate data sparsity~\cite{data_sparsity}. Existing knowledge-aware recommendation models can be roughly divided into three categories, i.e., embedding-based methods, path-based methods and propagation-based methods. 
For the embedding-based methods~\cite{CKE,embedding_based_DKN,embedding_based_RCF}, they primarily adopt the knowledge graph embedding (KGE) technique, e.g., TransR~\cite{TransR} and TransD~\cite{TransD}, to learn user and item representations by a transition constraint. 
For the path-based methods~\cite{PER,meta_path_v3}, they exploit the connectivity patterns among items guided by meta-paths to provide external semantic dependence to make recommendations.
For the propagation-based methods~\cite{KGAT,KGIN,MCCLK,KGIN}, they mainly perform recursive message propagation over KG to refine the user and item representations for making accurate recommendations. 
However, the maximum propagation depth of these models is usually limited to three to avoid the explosive increase of interacted neighbors~\cite{greater_than_2}, which hinders the ability of the models to explore high-order relatedness among nodes.
Differently, HiCON explicitly derives low- and high-order semantic relatedness by two message aggregation components linked in cascade, helping the model fully exploit structural and semantic information of KG.

\noindent
\textbf{Contrastive Learning for Recommendation.}
Contrastive learning aims to learn better user and item representations by pulling positive pairs closer and pushing negative pairs away~\cite{sim_cl}.
A great number of recommendation tasks, such as social recommendation~\cite{SEPT} and knowledge-aware recommendation~\cite{MCCLK,KGIC,KGCL}, benefit from this mechanism to improve recommendation performance. 
For example, for the social recommendation, SEPT~\cite{SEPT} exploits social relations between users to generate two complementary views based on tri-training for contrastive learning. 
As for knowledge-aware recommendation, MCCLK~\cite{MCCLK} performs a multi-level cross-view contrastive learning based on the local and global views of knowledge graph to learn effective user and item representations. 
However, existing knowledge-aware models perform contrastive learning based on vanilla GNNs equipped with limited propagation layers, which have  defects in capturing deeper semantic information. 
% \redColor{
On the contrary, we propose two message aggregation components linked hierarchically to encode shallow and deep semantic representations, which are treated as positive pairs for contrastive learning. 

\section{Preliminaries}
In this section, we introduce some basic concepts and the knowledge-aware recommendation task in this paper.

\noindent
\textbf{Bipartite Graph.} Let $\mathcal{U} = \{u_1, u_2, \cdots, u_m\}$  and $\mathcal{I} = \{v_1, v_2, \cdots, v_n\}$ be the sets of users and items in the recommendation system, where $m$ and $n$ are the numbers of users and items. 
We form a bipartite graph $\mathcal{G}_r$ from user-item interactions where its edges mean that users interact with items. 

\noindent
\textbf{Knowledge Graph.} Apart from the bipartite graph $\mathcal{G}_r$, we also provide a knowledge graph $\mathcal{G}_k$, and its formal definition is $\{(h, r, t) | h,t \in \mathcal{E}, r \in \mathcal{R} \}$ where $h$, $r$ and $t$ are the head, relation and tail of triplet $(h, r, t)$ that is the basic unit in knowledge graph. For example, (Brad Pitt, ActorOf, Fight Club) represents the fact that Brad Pitt is an actor in the movie Fight Club. Note that $\mathcal{R}$ contains bidirectional relations between nodes, such as ActorOf and its reverse relation ActedBy, to fully reveal the facts in KG. Besides, there exists an intersection between item set $\mathcal{I}$ and entity set $\mathcal{E}$, enabling the recommendation model to incorporate knowledge graph. 

\noindent
\textbf{Unified Graph.}
Following the collaborative knowledge graph in KGAT~\cite{KGAT}, we merge the bipartite graph $\mathcal{G}_r$ and knowledge graph $\mathcal{G}_k$ to construct a unified graph $\mathcal{G}_c = \{(h, r, t) | h,t \in \mathcal{E}', r \in \mathcal{R}' \}$, where $\mathcal{E}'=\mathcal{E} \cup \mathcal{U} \cup \mathcal{I}$ and $\mathcal{R}'=\mathcal{R} \cup \{r_{ui}\}$. $r_{ui}$ is the user-item interaction relation. 

\begin{figure}[t]
    \centering
    \includegraphics[width=0.48\textwidth]{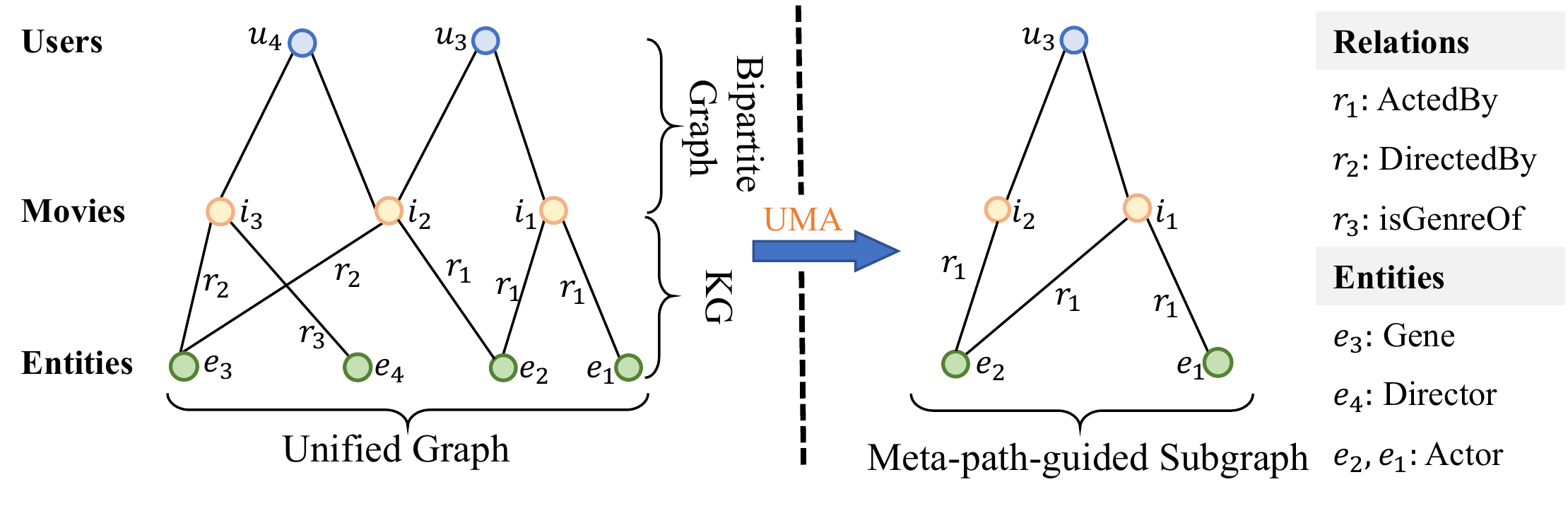}
    \vspace{-0.25in}
    \caption{An example of generating a meta-path-guided subgraph from a unified graph where UMA is the meta-path User-Movie-Actor. }
    \label{Fig:preliminaries}
    \vspace{-0.2in}
\end{figure}

\noindent
\textbf{Meta-path-guided Subgraph.} \label{sec:subgraph} 
Let $\mathcal{M} = \{m_1,\cdots, m_k\}$ be the set of k meta-paths.
A meta-path $m \in \mathcal{M} $ is denoted as a path in the form of $m = A_1\stackrel{R_1}{\longrightarrow}A_2 \cdots \stackrel{R_l}{\longrightarrow}A_{l+1}$ (abbreviated to $A_1A_2\cdots A_{l+1}$)~\cite{meta-path}.
It describes a composite relation $R = R_1 \circ  R_2 \cdots  R_l$ between two nodes with type $A_1$ and $A_{l+1}$ where $\circ$ is the composition operator on relations.
For example, Movie-Actor-Movie (MAM) demonstrates the semantic ``co-actor relationships between movies". 
Then, a meta-path-guided subgraph is defined as the combination of all path instances derived by the given meta-path. 
For example, given a unified graph and a meta-path User-Movie-Actor (UMA), as shown in Figure~\ref{Fig:preliminaries}, we conduct a subgraph including the path instances $u_3$-$i_2$-$e_2$, $u_3$-$i_1$-$e_2$ and $u_3$-$i_1$-$e_1$.

\noindent
\textbf{knowledge-aware Recommendation Task.} Given a unified graph $\mathcal{G}_c$, the goal is to recommend items to target users based on the relevance of their representations derived from $\mathcal{G}_c$. 

\begin{figure*}[t]
  \centering
 \includegraphics[width=1.0\linewidth]{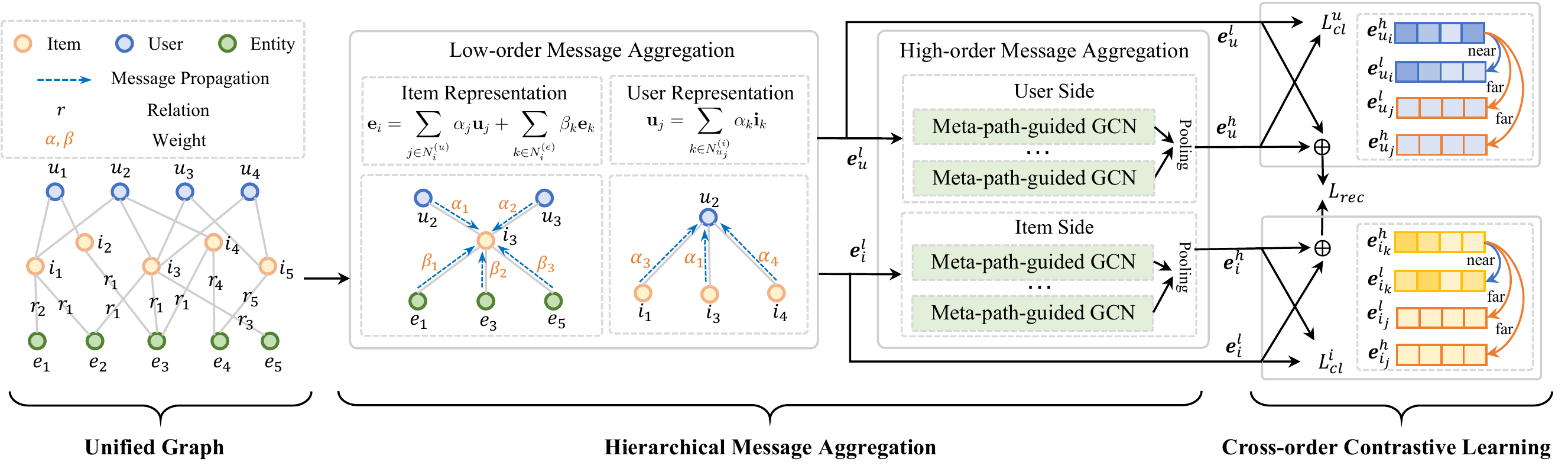}
  \vspace{-0.2in}
  \caption{The framework of our proposed HiCON model.}
  \vspace{-0.18in}
  \label{Fig:overall_framework}
\end{figure*}

\section{Method}

\subsection{Overall Framework}
The overall framework of HiCON is shown in Figure~\ref{Fig:overall_framework}. 
A hierarchical message aggregation module, consisting of two parts linked in a cascaded manner, learns low- and high-level node representations over the unified graph.
The former part aggregates the information of nodes' local neighbors by a limited number of common propagation layers.
The latter part reveals the valuable high-order semantic relations by employing graph convolutional networks over multiple subgraphs generated by the guidance of well-designed meta-paths.
Note that both parts help the model avoid propagating a vast number of useless neighbors to the central node.
Next, to further alleviate the over-smoothing issue by learning more discriminative node representations, we employ a cross-order contrastive learning module by contrasting low- and high-level representations of the same node.

\subsection{Hierarchical Message Aggregation}
Most existing GNN-based knowledge-aware methods, such as KGAT~\cite{KGAT} and KGCN~\cite{KGCN}, encode shallow (low-level) and deep (high-level) semantic information by a simple and naive recursive propagation mechanism. 
However, as the number of propagation layers increases, many central nodes are aware of more unrelated neighbors. This may make the model encounter the over-smoothing issue caused by bringing in a lot of noise information~\cite{hierarical_paper}. 
To alleviate this issue, we decompose the semantic representation learning into two message aggregation components combined in a hierarchical manner.

\noindent
\textbf{Low-order Message Aggregation.} 
The unified graph covers not only the user interaction behaviors about items, but also the item-wise semantic relatedness in knowledge graph.
To fully derive the shallow semantic information from the unified graph, 
we propose a dual propagation layer consisting of two embedding propagation layers for the bipartite graph and knowledge graph, respectively.

For the bipartite graph, we utilize a light embedding propagation layer to aggregate the neighbors based on the topological structure of the bipartite graph.
Formally, the definition can be summarized as follow:
\begin{gather}
  \alpha_{i,j} = \frac{1}{\sqrt{|\mathcal{N}_{u_i}^{(v)}|}\sqrt{|\mathcal{N}_{v_i}^{(u)}|}}, \ \ 
 % \hat{\textbf{e}}_{v_{i}}^{(k+1)} 
 \textbf{h}_{v_{i}}^{(k+1)} = \sum_{j \in \mathcal{N}_{v_i}^{(u)}}  \alpha_{i,j} \textbf{e}_{u_{j}}^{(k)}, \label{eq:LGC_1}
\end{gather}
where $\mathcal{N}_{u_i}^{(v)}$ is the set of items interacted by user $u_i$, and 
% $\hat{\textbf{e}}_{v_{i}}^{(k+1)}$ 
$\textbf{h}_{v_{i}}^{(k+1)}$
is the user-side representation of item $v_i$ in the $k$+1-th layer.
For knowledge graph, due to there are various types of relations among nodes (e.g., ActedBy and DirectedBy), we propose a relation-aware embedding propagation layer relying on the attention mechanism 
to exploit the structural and semantic information of KG.
Concretely, given an item $v_i$, and a triplet set $\mathcal{N}_{v_i}^{(e)}= \{ (v_i,r,t)|(v_i,r,t) \in \mathcal{G}_k\}$ with the item $v_i$ as the head entity, we perform relation-aware attention to aggregate the neighbors within the set:
$
\mathbf{g}_{v_i}^{(k+1)}
= \sum_{(v_i,r,t)\in \mathcal{N}_{v_i}^{(e)}} \beta_{v_i,t}^{(k)} (\mathbf{e}^{(k)}_{r} \odot \mathbf{e}^{(k)}_{t})
\label{eq:add}
$
where $\odot$ is the element-wise multiplication. 
$\mathbf{g}_{v_i}^{(k+1)}$ is the entity-side representation for item $v_i$ in the $k$+1-th layer, and $\mathbf{e}^{(k)}_{t}$ and $\mathbf{e}^{(k)}_{r}$ are the representations for the tail entity $t$ and relation $r$ in the $k$-th layer. The attention weight $\beta_{v_i,t}^{(k)}$ is normalized by a softmax function based on the importance coefficient $\pi_{v_i,t}^{(k)}$:
\begin{align}
\beta_{v_i,t}^{(k)} = \frac{\exp(\pi_{v_i,t}^{(k)})}{\sum_{(v_i,r,t)\in \mathcal{N}_{v_i}^{(e)}} \exp({\pi_{v_i,t}^{(k)}})}, \ 
\pi_{v_i,t}^{(k)} = \textbf{s}_{v_i}^T \textbf{s}_{t},
\end{align}
where $\textbf{s}_{v_i} = \mathbf{e}^{(k)}_{v_i} \odot \mathbf{e}^{(k)}_r/\| \mathbf{e}^{(k)}_{v_i} \odot \mathbf{e}^{(k)}_r \|$ considers the influence of relation $r$.  $\| \cdot \|$ is the $\mathit{l}_2$-norm, and similarly for $\textbf{s}_{t}$. 

Since the items are regarded as the bridge connecting the bipartite graph and knowledge graph, we merge both collaborative signals and semantic information of KG to further refine the item representations. 
Concretely, given the user-side representation $\textbf{h}_{v_{i}}^{(k+1)}$ and entity-side representation $\textbf{g}_{v_{i}}^{(k+1)}$ of item $v_i$, we combine these two representations linearly with the sum-pooling operation to acquire the refined item representation $\textbf{e}_{v_{i}}^{(k+1)}$ in the $k$+1 layer: $  \textbf{e}_{v_{i}}^{(k+1)} = \textbf{h}_{v_{i}}^{(k+1)} + \textbf{g}_{v_{i}}^{(k+1)}$.
Finally, to fully exploit the shallow semantic relatedness, we stack two dual propagation layers and sum the output of layers up as the low-level item representations: $\textbf{e}_{v_{i}} = \mathbf{e}_{v_i}^{(0)} + \mathbf{e}_{v_i}^{(1)} + \mathbf{e}_{v_i}^{(2)}$, and similarly for the low-level user representations $\textbf{e}_u$.

\begin{figure}[t]
  \centering
 \includegraphics[width=1.0\linewidth]{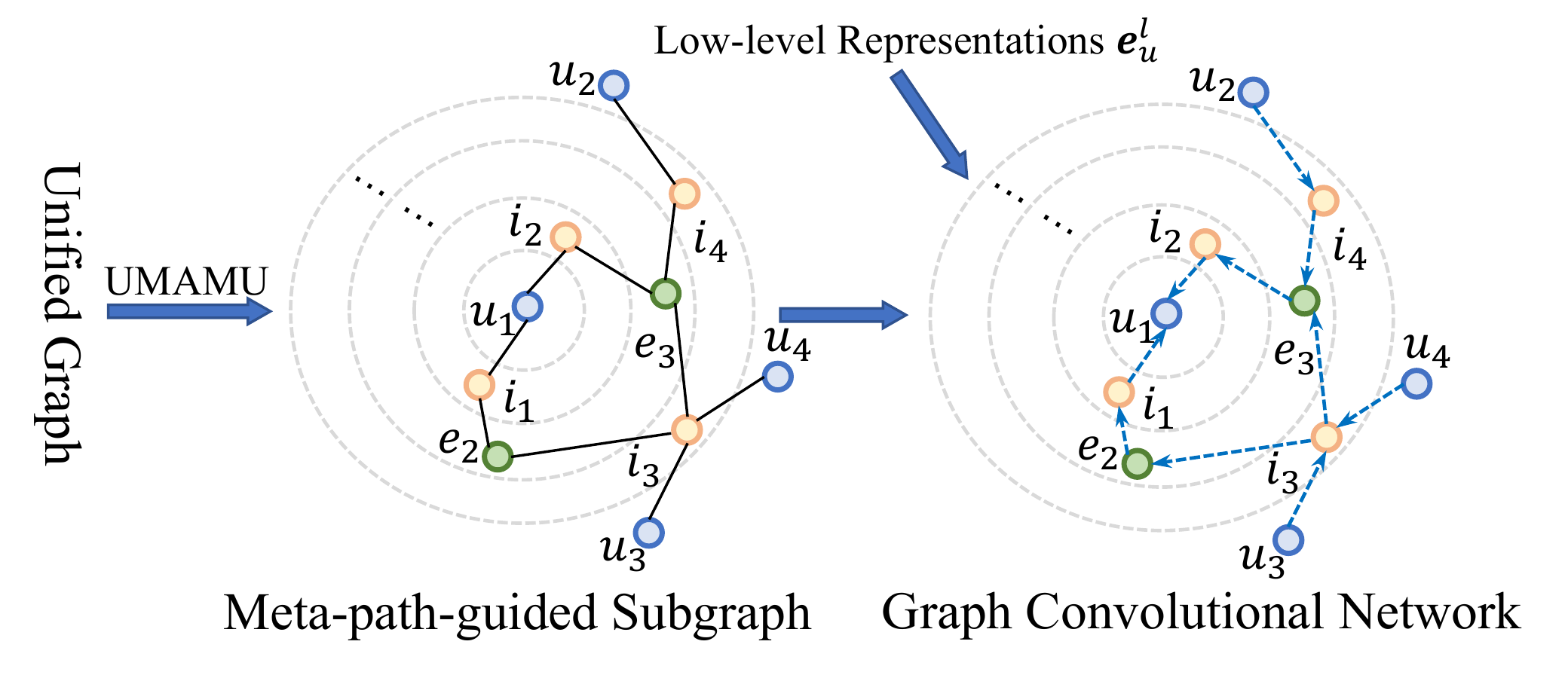}
 \vspace{-0.25in}
  \caption{An example of a meta-path-guided GCN where UMAMU denotes the meta-path User-Movie-Actor-Movie-User.
  The concentric circles denote different hops of the user.}
  \vspace{-0.15in}
  \label{Fig:high_propagation}
\end{figure}

\noindent
\textbf{High-order Message Aggregation:} 
In this section, we perform the meta-path-guided message aggregation to derive useful high-order relatedness (shown in Figure~\ref{Fig:high_propagation}). 
Concretely, given a meta-path $m$ and the unified graph $\mathcal{G}_c$, we collect all path instances derived by the meta-path from $\mathcal{G}_c$ to construct a subgraph $\mathcal{G}_s^m$ (more details on the construction of this subgraph refer to Section~\ref{sec:subgraph}). 
To capture the high-order semantic relatedness from $\mathcal{G}_s^m$, we use a graph convolutional network (GCN) where its single graph convolutional layer employs a nonlinear transformation after the combination of central users and their neighbors: 
\begin{align}
\mathbf{z}^{(k+1)}_{m,u_i} = \sigma \bigg (\sum_{j \in \mathcal{N}^{(m)}_{u_i}} \frac{1}{\sqrt{|\mathcal{N}^{(m)}_{u_i}|} \sqrt{|\mathcal{N}^{(m)}_{j}|}} \mathbf{z}^{(k)}_{m,j} \mathbf{W}^{(k)}_m\bigg ),
\end{align}
where $\sigma(\cdot)$ is a non-linear activation function, and $\mathbf{W}^{(k)}_m$ is a weight matrix in the $k$-th layer. 
$\mathcal{N}^{(m)}_{u_i}$ is the set of neighbors for the user $u_i$ in $\mathcal{G}_s^m$.
$\textbf{z}_{m,u_i}^{(k+1)}$ is the representation of user $u_i$ encoded from $\mathcal{G}_s^m$ in the $k+1$-th layer. 
We select the output of the last layer of GCN as the final high-level representations $\textbf{z}_{v_i}^m$ rather than concatenating or summing the output of all layers to reduce the noise disturbance: $\textbf{z}_{u_i}^m = \mathbf{z}^{(l)}_{m,u_i}$, 
where $l$ is the layer number in GCN. 
To effectively exploit diverse high-order relatedness among nodes, we develop multiple meta-paths for constructing corresponding subgraphs, and then utilize independent GCNs to encode multiple high-level user representations. Finally, a mean-pooling operation is applied to combine these representations: $\textbf{z}_{u_{i}} = \sum_{m\in \mathcal{M}_u}\mathbf{z}_{u_i}^m $, 
where $\mathcal{M}_u$ is the set of meta-paths on the user side, and similarly for the high-level item representations $\textbf{z}_v$.

\subsection{Cross-Order Contrastive Learning}
In this section, we propose a cross-order contrastive learning module to learn discriminative node representations by distinguishing positive pairs from negative ones.
Concretely, considering the low-level item vectors $\mathbf{e}_{v}$ and high-order item vectors $\mathbf{z}_{v}$ as positive pairs, we present an item-side contrastive learning loss to minimize the distance between the positive pairs:  
$
\mathcal{L}_{cl}^{I}= - \sum\limits_{v_i \in \mathcal{I}} \log\frac{\exp({(\textbf{e}_{v_i} \cdot \textbf{z}_{v_i})/\tau})}{\sum\limits_{v_j \in \mathcal{C}_{v_i} \cup \{ v_i\}} \exp({(\textbf{e}_{v_i} \cdot \textbf{h}_{z_j})/\tau}}),
$
where $\mathcal{C}_{v_i}$ is a negative sample set, consisting of items in the batch except item $v_i$, and $\tau$ is the temperature parameter, which is set to 0.6. The user-side loss $\mathcal{L}_{cl}^{U}$ is calculated similarly. 
After that, we linearly combine the item-side loss $\mathcal{L}_{cl}^{I}$ and user-side loss $\mathcal{L}_{cl}^{U}$ to get the final object for cross-order contrastive learning:
$
\mathcal{L}_{cl} = \mathcal{L}_{cl}^{U} + \mathcal{L}_{cl}^{I}.
$
\subsection{Optimization and Prediction}
To optimize HiCON, we adopt a pairwise Bayesian personalized ranking (BPR) loss $\mathcal{L}_{bpr}$ equipped with the aforementioned contrastive loss $\mathcal{L}_{cl}$ as the final loss $\mathcal{L}_{HiCON}$ by the weighted sum operation:
\begin{gather}
\mathcal{L}_{HiCON} = \mathcal{L}_{bpr}+ \lambda \mathcal{L}_{cl}, 
\end{gather}
where $\lambda$ is a hyper-parameter to control the balance between the losses. 
$\mathcal{L}_{bpr}$ imposes users to have higher predicted scores with interacted items than non-interacted ones, which is defined as: 
$ 
\mathcal{L}_{bpr} = \sum_{u \in \mathcal{U}} \sum_{i \in \mathcal{N}_u} \sum_{i' \not \in \mathcal{N}_u} - \log \sigma(\hat{y}_{u,i} - \hat{y}_{u,i'}),
$
where $\mathcal{N}_u$ denotes the item neighbors of user $u$ in the bipartite graph. 
The probability $\hat{y}_{u,i} = [\mathbf{e}_u \mathbin\Vert \mathbf{z}_u]^T [\mathbf{e}_i \mathbin\Vert \mathbf{z}_i]$ is calculated by the dot product between the concatenated user and item representations where $\mathbin\Vert$ denotes the concatenate operation, which is also used to predict user's preferred items during the inference phase. 

\section{Experiment}

\subsection{Experimental Settings}
\noindent
\textbf{Datasets.} 
We conduct experiments on three widely used benchmark datasets for recommendation, i.e., LastFM\footnote{ https://grouplens.org/datasets/hetrec- 2011/}, Book-Crossing\footnote{http://www2.informatik.uni- freiburg.de/~cziegler/BX/} and MovieLens-1M\footnote{https://grouplens.org/datasets/movielens/1m/},
which are collected from the music, book, and movie domains, respectively.
The details of the datasets are included in Appendix A.1 (including statistics and data preprocessing).

\noindent
\textbf{Baseline Methods.} To verify the effectiveness of HiCON, we select two representative groups of recommendation models as baseline methods: collaborative filtering-based group (BPRMF~\cite{bprmf}, LightGCN~\cite{lightgcn}) and knowledge-aware group.
In the knowledge-aware group, there are three classic types of methods: an embedding-based method (CKE~\cite{CKE}), a path-based method (PER~\cite{PER}), and propagation-based methods (RippleNet~\cite{Ripplenet}, KGCN~\cite{KGCN}, KGNN-LS~\cite{KGNN-LS}, KGAT~\cite{KGAT}, CKAN~\cite{CKAN}, KGIN~\cite{KGIN}, MCCLK~\cite{MCCLK}, KGIC~\cite{KGIC}). 
For the implementation of baselines, please refer to Appendix A.2.

\noindent
\textbf{Evaluation Metrics \& Parameter Settings.} We use two classical metrics \textit{Area Under Curve} (AUC) and F1, which are widely used in click-through rate (CTR) prediction, to evaluate all models. 
We tailor multiple meta-paths for each relation on both item and user sides. 
Specifically, the item-side meta-path set $\mathcal{M}_v$ includes Item-User-Item-Entity-Item (IUIEI), Item-Entity-Item-User-Item (IEIUI), and Item-User-Item-User-Item (IUIUI). 
The user-side set $\mathcal{M}_u$ contains User-Item-User-Item-User (UIUIU) and User-Item-Entity-Item-User (UIEIU). 
The weight $\lambda$ for controlling the balance of losses is set to 0.01 on MovieLen-1M, and 0.05 on the remaining datasets. 
\subsection{Experimental Results}

\noindent
\textbf{Overall Performance Comparison.}
\begin{table}[t]
    \caption{Performance comparison of HiCON and baselines. The best and second-best results are indicated by bold fonts and underlines, and * indicates statistically significant improvements over the best baseline (t-test with p$<$0.01).}
    \centering
    \scalebox{0.75}{
    \begin{tabular}{cllllll}
    	\hline
    	\multirow{2}{*}{Model} & \multicolumn{2}{c}{Book-Crossing} & \multicolumn{2}{c}{MovieLens-1M} & \multicolumn{2}{c}{Last.FM} \\
        & \multicolumn{1}{c}{\textit{AUC}} & \multicolumn{1}{c}{\textit{F1}} & \multicolumn{1}{c}{\textit{AUC}} & \multicolumn{1}{c}{\textit{F1}} & \multicolumn{1}{c}{\textit{AUC}} & \multicolumn{1}{c}{\textit{F1}} \\
        \hline
BPRMF & 0.6583 & 0.6117 & 0.8920 & 0.7921 & 0.7563 & 0.7010 \\
LightGCN & 0.6134 & 0.6469 & 0.8800 & 0.8091 & 0.8300 & 0.7439  \\ \hline
CKE   & 0.6759 & 0.6235 & 0.9065 & 0.8024 & 0.7471 & 0.6740 \\ \hline
PER   & 0.6048 & 0.5726 & 0.7124 & 0.6670 & 0.6414 & 0.6033 \\ \hline
RippleNet  & 0.7211 & 0.6472& 0.9190 & 0.8422 & 0.7762 & 0.7025 \\ 
KGCN& 0.6841 & 0.6313 & 0.9090 & 0.8366& 0.8027& 0.7086 \\
KGNN-LS  & 0.6762 & 0.6314 & 0.9140& 0.8410 & 0.8052 & 0.7224\\
KGAT & 0.7314 & 0.6544 & 0.9140 & 0.8440& 0.8293& 0.7424        \\
CKAN & 0.7420 & 0.6671 & 0.9082 & 0.8410 & 0.8418 &0.7592 \\
KGIN & 0.7273 & 0.6614 & 0.9190 & 0.8441 & 0.8486 & 0.7602 \\
MCCLK  & 0.7625 & 0.6777 & \underline{0.9351} & \underline{0.8631} & \underline{0.8763} & \underline{0.8008} \\ 
KGIC & \underline{0.7749} & \underline{0.6812} & 0.9252 & 0.8559 & 0.8592 & 0.7753 \\
\hline
HiCON & \textbf{0.8045*} & \textbf{0.7169*} & \textbf{0.9410*} & \textbf{0.8718*} & \textbf{0.9045*} & \textbf{0.8186*}  \\
\hline
\end{tabular}}
\label{tab:performance}
\end{table}
Table~\ref{tab:performance} shows the results of HiCON and baselines on all datasets.
We find that HiCON achieves the best performance compared with all baselines.
This is because the hierarchical message aggregation effectively derives different-level semantic information from the local semantic relatedness and meta-path-guided high-order relatedness. 
Besides, the cross-order contrastive learning module can improve the self-discrimination of node representations. 
We also find that, due to the rich structural and semantic information in KG, most knowledge-aware recommendation models outperform collaborative filtering-based methods without merging KG (BPRMF and LightGCN) by a large margin. 
In addition, the superior performance of most propagation-based methods over embedding-based (CKE) or path-based ones (PER), which may attribute to recursive propagation mechanisms that mine deeper semantic relatedness.

\noindent
\textbf{Ablation Study.}
\begin{figure}[t]
  \centering
 \includegraphics[width=1.0\linewidth]{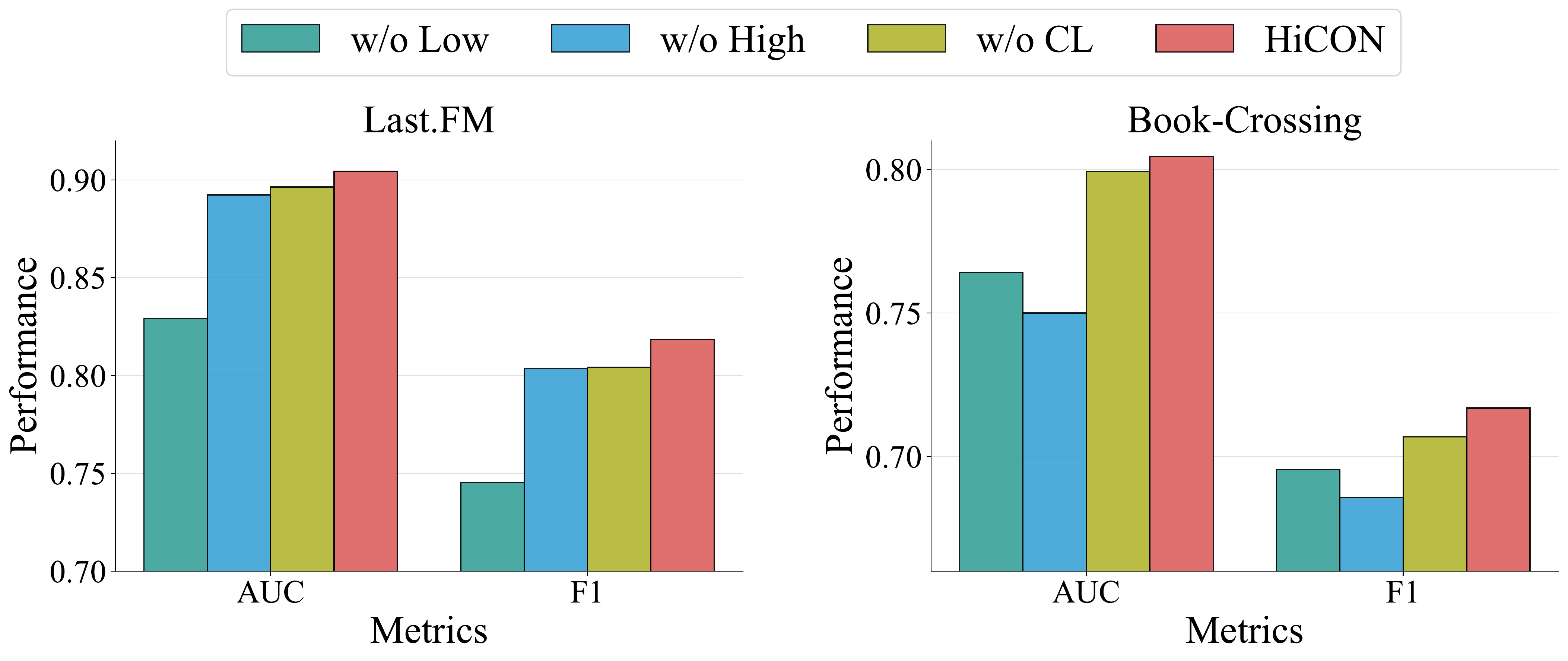}
  \caption{The performance of HiCON and its variants on the LastFM and Book-Crossing datasets.}
  \label{Fig:ablation_study}
  % \vspace{-0.1in}
\end{figure}
To analyze the effectiveness of key components in HiCON, we compare the performance of HiCON with its three variants: 
1) w/o Low removes the low-order message aggregation of HiCON; 
2) w/o High is the variant of HiCON without the high-order message aggregation;
3) w/o CL wipes out the cross-order contrastive learning module.
The results are summarized in Figure~\ref{Fig:ablation_study}. 
First, after removing low- and high-order message aggregations, the performance of HiCON significantly decreases.
It shows that both aggregations derive useful external semantic knowledge for learning better user and item representations. 
Second, hierarchically combining two message aggregations can also improve performance, indicating the low- and high-level representations are mutually enhanced to further provide more precise recommendations.
Third, contrastive learning can boost the performance, indicating the discriminative user and item representations are helpful for making accurate recommendation.

In addition, we visualize the feature distribution of user representations in Appendix A.3 to indicate the ability of each module of HiCON to alleviate over-smoothing.
The results show that the hierarchical message aggregation
and cross-order contrastive learning modules both derive more uniform feature distributions than the variant based on recursive message propagation, indicating both modules help HiCON learn more diverse user preferences to alleviate over-smoothing. 

\begin{figure}[t]
  \centering
 \includegraphics[width=1.0\linewidth]{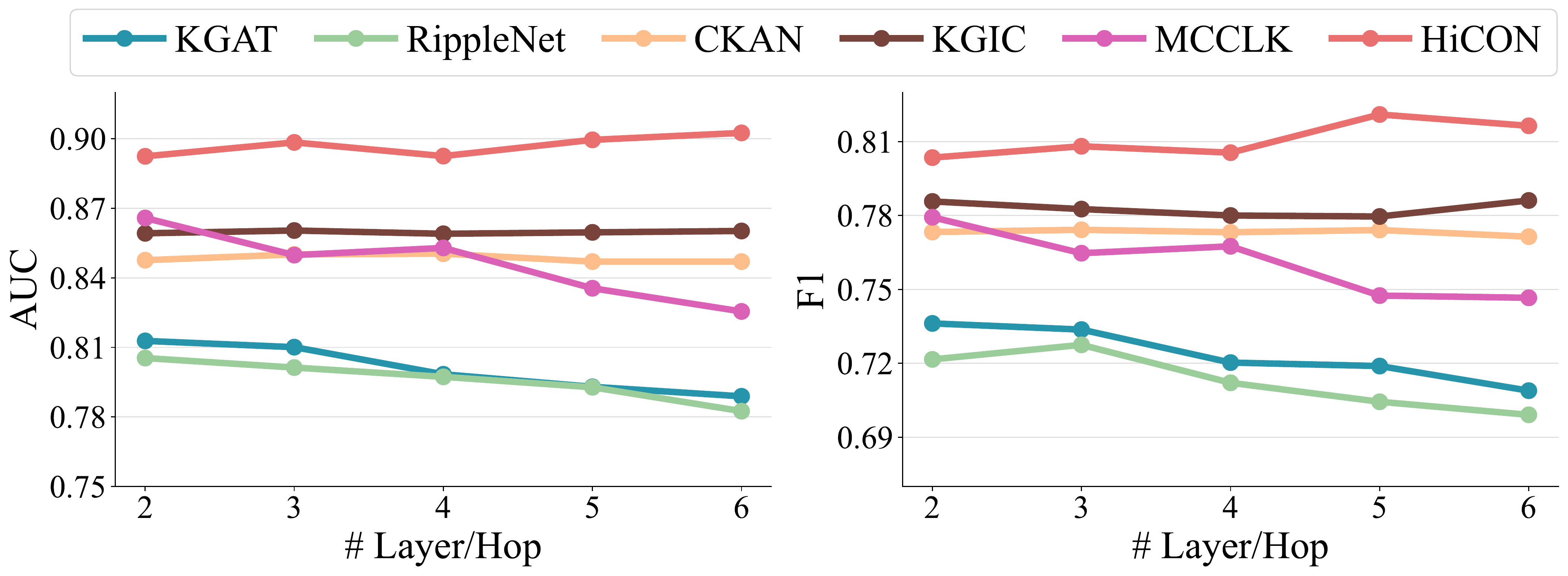}
 \vspace{-0.1in}
  \caption{The performance of HiCON and baselines at different layers (hops) on the LastFM dataset.}
  \label{Fig:performance_hop_number}
\end{figure}

\noindent
\textbf{Performance w.r.t Hop (Layer) Number.}
Here we compare the performance of HiCON with other baselines under different layers (hops). 
The results are shown in Figure~\ref{Fig:performance_hop_number}. 
We can observe some interesting findings. 
First, the performance of some baselines (i.e., KGAT, RippleNet and MCCLK) gradually declines along with the increase of the layer (hop) number. 
This is because they are aware of exponentially growing neighbors during recursive message propagation, which is harmful to personalized node representation learning. 
On the contrary, CKAN and KGIC achieve stable performance at different layers (hops).
This is because they select a fixed number of neighbors by random sampling.
At last, compared with all baselines, HiCON achieves the best performance at all layers (hops), indicating the effectiveness of HiCON in capturing both low and high-order semantic relatedness and alleviating the over-smoothing issue.

\noindent
\textbf{Additional Experiments.} We perform experiments in Appendix A.4 and A.5 to indicate the effectiveness of HiCON in modeling high-order semantic relatedness and at various interaction sparsity levels, respectively. Besides, we also analyze the impact of hyperparameter $\gamma$ in Appendix A.6, and find that over-focusing the contrastive loss $\mathcal{L}_{cl}$ may hinder the model optimization process.

\section{Conclusion}
In this paper, we propose a hierarchical and contrastive representation learning framework 
to alleviate the over-smoothing issue for effectively absorbing the structural and semantic information of KG. 
It relieves this issue by avoiding the exponential expansion of neighbors and enhancing the self-discrimination of node representations.
We first propose a hierarchical message aggregation mechanism to explore different-level semantic relatedness via aggregating local neighbors and meta-path-guided high-order neighbors.
Besides, we further perform cross-order contrastive learning via maximizing the consistency between low- and high-level views.
Extensive experiments on three datasets show the superiority of HiCON over state-of-the-art methods and the ability of alleviating over-smoothing.

% \small
% \footnotesize
\bibliographystyle{IEEEbib}
\bibliography{./icme}

\appendix
\section{Appendix}
\subsection{Datasets}
Here we introduce the details and preprocessing of datasets.
Last.FM\footnote{ https://grouplens.org/datasets/hetrec- 2011/} is collected from Last.fm music platform where musical tracks are regarded as items. 
Book-Crossing\footnote{http://www2.informatik.uni- freiburg.de/~cziegler/BX/} is a dataset consisting of ratings about books in the Book-Crossing community. 
MovieLens-1M\footnote{https://grouplens.org/datasets/movielens/1m/} is a classical dataset collecting movie ratings from the MovieLens website for movie recommendation.  
Following previous works~\cite{Ripplenet,KGIC}, we omit the user-item pairs whose ratings are less than the pre-defined threshold (it is set to 4 for MovieLens-1M and 1 for the remaining datasets) and treat the other pairs as positive samples with label 1. 
The negative instances are randomly sampled from unobserved items with the same number of positive ones for each user. 
As for the knowledge graph construction, we follow RippleNet~\cite{Ripplenet} to select item-related triplets from Microsoft Satori\footnote{https://searchengineland.com/library/bing/bing- satori} for each dataset.
The statistics of these datasets are summarized in Table~\ref{Tab:dataset}.
\begin{table}[t]
\centering
\caption{Statistics of the LastFM, Book-Crossing, and MovieLens-1M datasets.}
\scalebox{0.75}{
\begin{tabular}{cl|ccc}
\hline
\multicolumn{1}{l}{}                                                                &                 &
\multicolumn{1}{l}{Last.FM} &
\multicolumn{1}{l}{Book-Crossing} & \multicolumn{1}{l}{MovieLens-1M}  \\ \hline
\multirow{3}{*}{\begin{tabular}[c]{@{}c@{}} Bipartite   \\ Interaction\end{tabular}} & \# users & 1,872 & 17,860 & 6,036                       \\
& \# items & 3,846 & 14,967 & 2,445                       \\
& \# interactions & 42,346 & 139,746 & 753,772                       \\ \hline
\multirow{3}{*}{\begin{tabular}[c]{@{}c@{}}Knowledge\\ Graph\end{tabular}} & \# entities & 9,366 & 77,903 & 182,011                        \\
& \# relations & 60 & 25 & 12                           \\
& \# triplets & 15,518 & 151,500 & 1,241,996                       \\ \hline
\end{tabular}}
\label{Tab:dataset}
\end{table}

\subsection{Implementation of Baselines}
Here we introduce the details of baselines used in this paper, which include two groups of representative methods: i.e., collaborative filtering-based methods (BPRMF, LightGCN) and knowledge-aware methods. In the knowledge-aware methods, we employ three groups of typical methods: i.e., embedding-based methods (CKE), path-based methods (PER), and propagation-based methods (RippleNet, KGCN, KGNN-LS, KGAT, CKAN, KGIN, MCCLK, KGIC):
\begin{itemize}
    \item \textit{BPRMF}~\cite{bprmf} is a representative matrix factorization-based method optimized by the Bayesian personalized ranking (BPR) object.
    \item \textit{LightGCN}~\cite{lightgcn} is a simplified graph convolutional network (GCN) by removing the feature transformation and non-linear activation. 
    \item \textit{CKE}~\cite{CKE} is an embedding-based recommendation method that exploits structural, textual and visual information under a Bayesian framework.
    \item \textit{PER}~\cite{PER} is a path-based method that extracts meta-path-aware features to improve the ability to capture the connectivity between users and items.
    \item \textit{RippleNet}~\cite{Ripplenet} is a classical propagation-based method by propagating the user preference along KG links to discover potential interests.
    \item \textit{KGCN}~\cite{KGCN} aggregates selected neighbors to learn both structural and semantic information of KG based on the graph convolutional network.
    \item \textit{KGNN-LS}~\cite{KGNN-LS} is a propagation-based method that equips GNNs with a label smoothness regularization mechanism to enrich node representations.
    \item \textit{KGAT}~\cite{KGAT} is a propagation-based method that recursively integrates neighbors over the unified graph with an attention strategy.
    \item \textit{CKAN}~\cite{CKAN} independently encodes collaborative information from the bipartite graph and semantic relationships from knowledge graphs relying on distinct message propagation strategies.
    \item \textit{KGIN}~\cite{KGIN} considers fine-grained user intents to profile user-item relationships, and uses relational paths to capture the long-range connectivity.
    \item \textit{MCCLK}~\cite{MCCLK} utilizes a multi-level contrastive learning mechanism based on local and global semantic views to enrich user and item representations. 
    \item \textit{KGIC}~\cite{KGIC} performs multi-level interactive contrastive learning based on layer-wise augmented views.
\end{itemize}

\subsection{Representation Visualization}
\begin{figure}[t]
  \centering
  \begin{subfigure}{0.15\textwidth}
    \centering
      \includegraphics[scale=0.12]{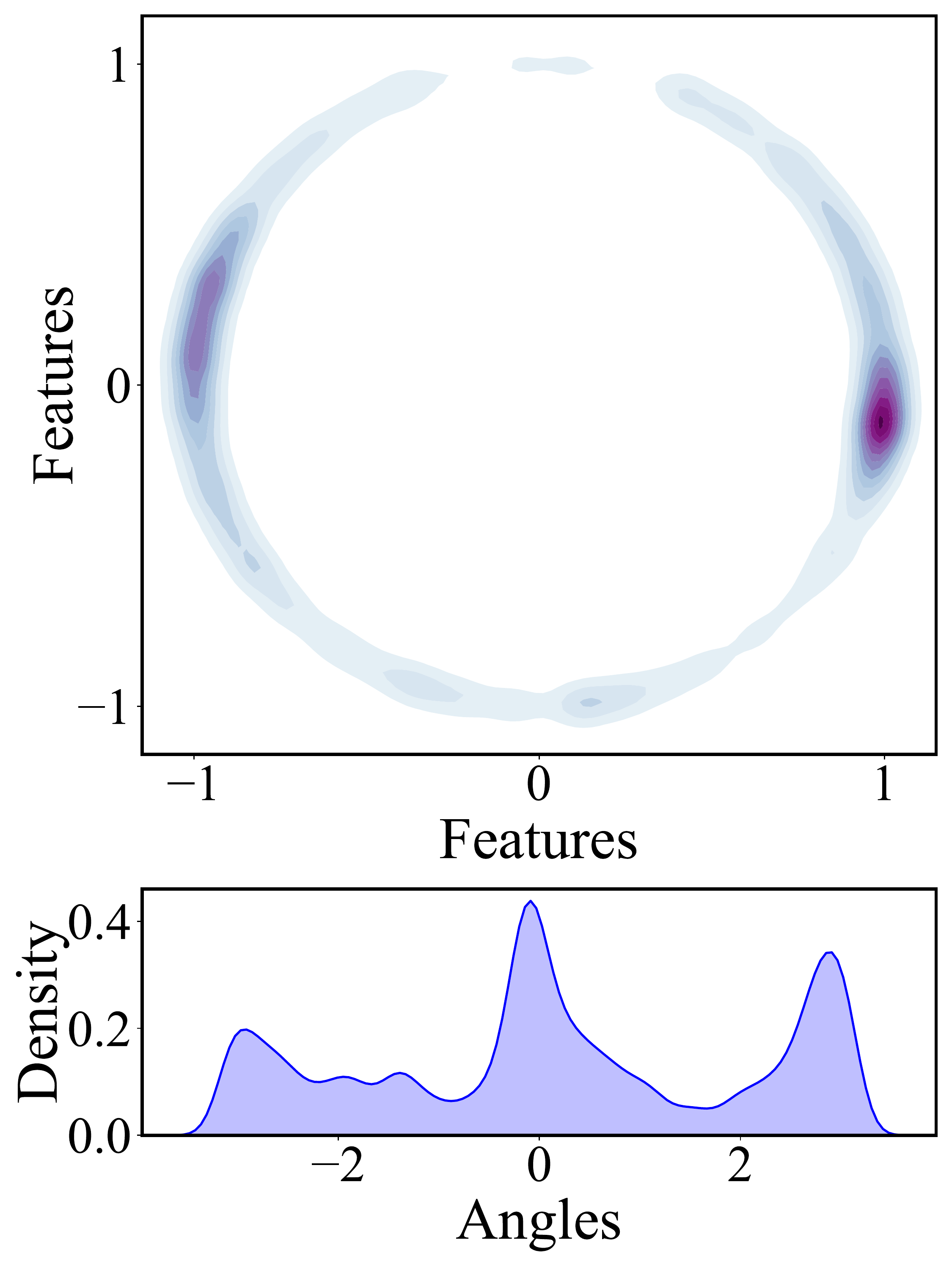}
    \caption{HiCON$_{w/o \ Hie}$}
    \label{Fig:oversmoothing}
  \end{subfigure}
  \begin{subfigure}{0.15\textwidth}
    \centering
      \includegraphics[scale=0.12]{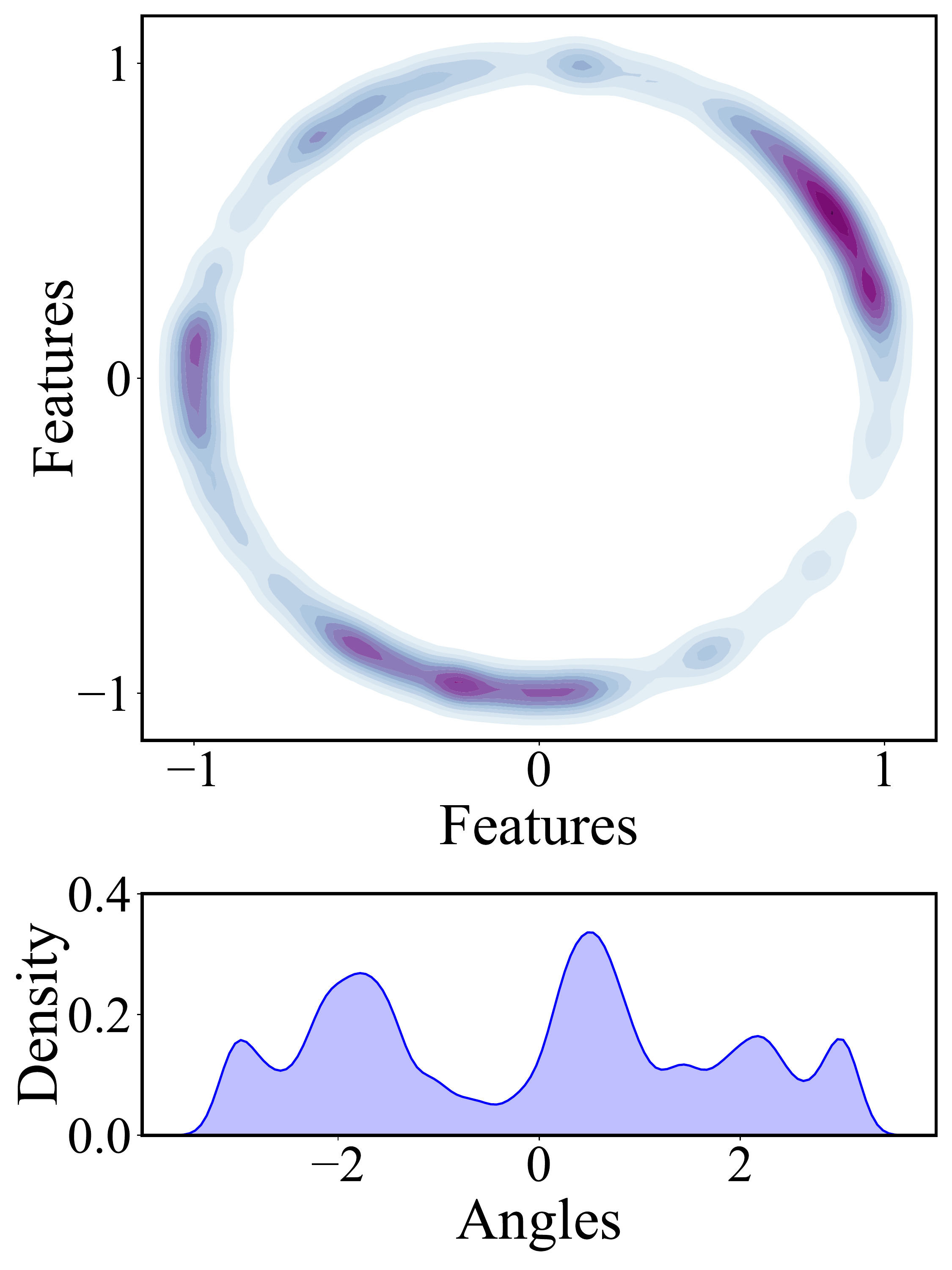}
    \caption{HiCON$_{w\ Hie}$}
    \label{Fig:high-order-issue}
  \end{subfigure}
  \begin{subfigure}{0.15\textwidth}
    \centering
      \includegraphics[scale=0.12]{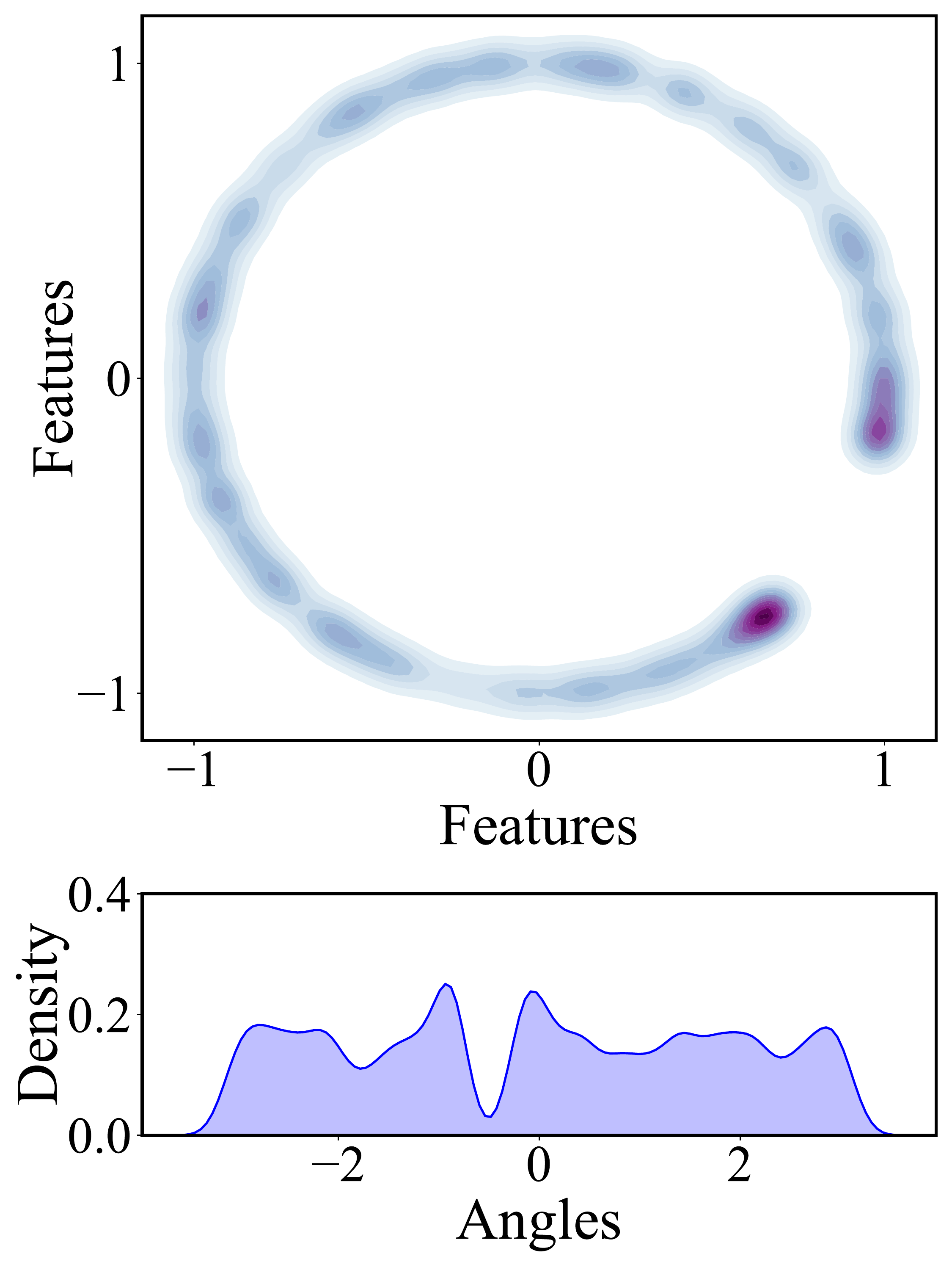}
    \caption{HiCON}
    \label{Fig:high-order-issue}
  \end{subfigure}
  \caption{The visualization of user representations derived by HiCON and its variants on the LastFM dataset. The feature distributions (the darker the color, the more points are distributed into the area~\protect\cite{sim_cl}.) and density estimations (the sharper the curve, the more points are clustered together.) of user representations are plotted in the unit hypersphere.}
  \vspace{-0.1in}
  \label{Fig:over_smoothing_visual}
\end{figure}

In this section, we conduct experiments to analyze the over-smoothing issue by visualizing the user representations learned by HiCON and its two variants: 
\begin{itemize}
    \item HiCON$_{w/o\ Hie}$ employs the recursive message propagation with the same receptive field of HiCON to replace the hierarchical message aggregation.
    \item HiCON$_{w\ Hie}$ merely utilizes the hierarchical message aggregation of HiCON and does not perform cross-order contrastive learning. 
\end{itemize}
Concretely, we first map the user representations into 2-dimensional normalized vectors by t-SNE~\cite{t-sne}. 
Then, we plot feature distributions with Gaussian kernel Density Estimation (KDE)~\cite{kde} in $\mathbb{R}^2$. 
To present the representations more intuitively, we plot density estimations on angles (i.e., $\arctan2(y, x)$ for each point $(x, y)$ in the unit hypersphere)~\cite{sim_cl}.
The visualization is shown in Figure~\ref{Fig:over_smoothing_visual}, from which we observe some findings.
First, we observe that the user representations derived from HiCON$_{w/o\ Hie}$ are mapped into a narrow area. This may be because the recursive message propagation forces the users to be aware of vast graph neighbors, causing the users to share similar representations in the latent space.  
Second, we observe a more uniform feature distribution and a more flat density curve when replacing the recursive message propagation with hierarchical message aggregation.
This may be because the hierarchical aggregation interacts with a collection of valuable neighbors rather than exponentially growing graph neighbors to reduce noise disturbance.
Third, compared with HiCON$_{w\ Hie}$, the distribution and the curve of HiCON respectively become more uniform and flat.
This may be because performing cross-order contrastive learning enhances the self-discrimination of user representations.

\subsection{Performance w.r.t High-order Semantic Relatedness Modeling}
\begin{figure}[t]
  \centering
 \includegraphics[width=1.0\linewidth]{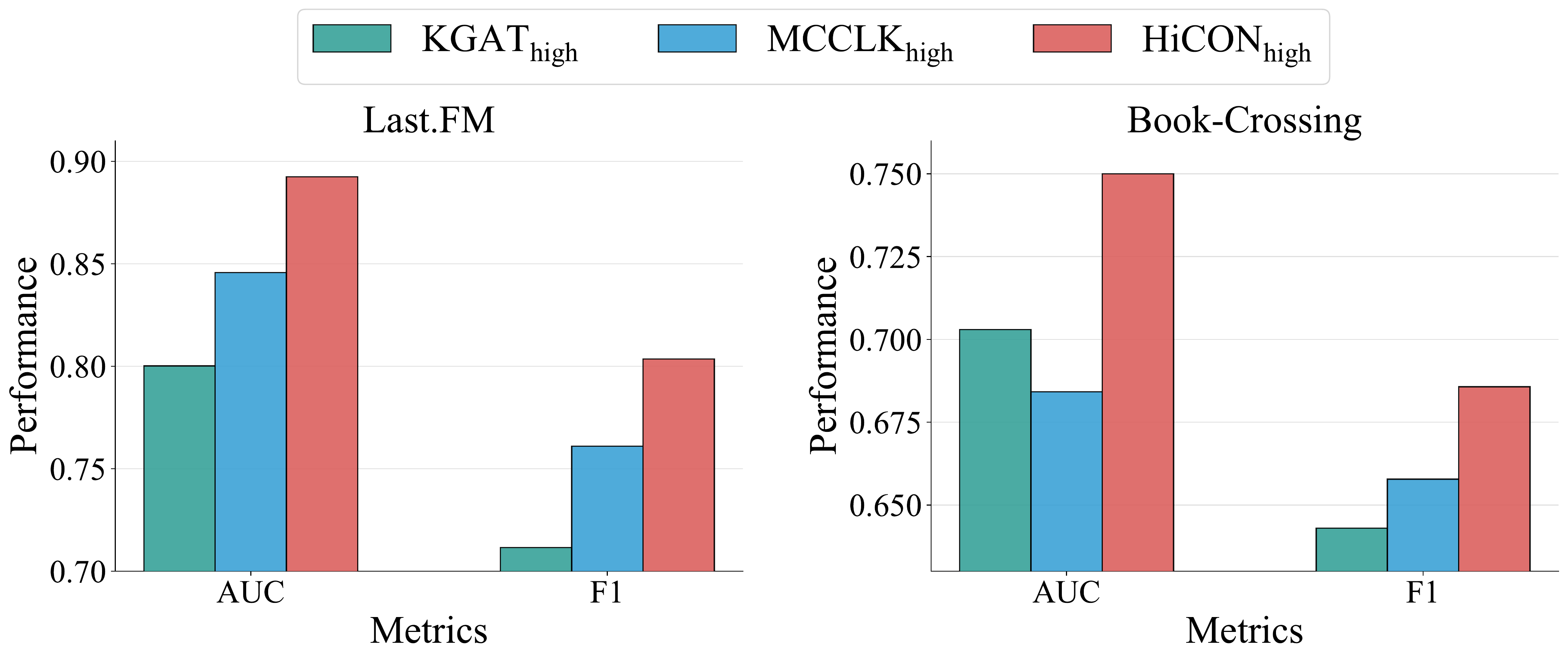}
  \caption{The performance of KGAT, MCCLK, and HiCON in modeling the high-order semantic relatedness on the LastFM and Book-Crossing datasets.}
  \label{Fig:high_rep_performance}
\end{figure}

In this section, we conduct experiments to validate the effectiveness of HiCON in modeling high-order semantic relatedness. 
To achieve this goal, we compare the performance of HiCON with two mainstream baselines, i.e., KGAT~\cite{KGAT} and MCCLK~\cite{MCCLK}. 
To successfully capture the high-order semantic relatedness over the unified graph, we set the number of layers (hops) of all models to six for a fair comparison.
More precisely, we conduct the following models: 
\begin{itemize}
    \item KGAT$_{high}$ uses the output of the last layer of KGAT as the final representations for recommendation;
    \item MCCLK$_{high}$ regards the output of the last layer of MCCKL as the final representations;
    \item HiCON$_{high}$ merely utilizes the output of the high-order message aggregation for recommendation. 
\end{itemize} 
The results of recommendation performance are shown in Figure~\ref{Fig:high_rep_performance}. 
We observe that HiCON$_{high}$ outperforms the other two models by a large margin, indicating that our proposed model effectively encodes high-order relatedness to derive high-quality deep semantic representations.
This may be because the hierarchical message propagation involved in HiCON selects and propagates a bundle of valuable neighbors to the central nodes rather than considering all exponentially increasing neighbors to reduce noise disturbance.

\begin{figure}[t]
  \centering
 \includegraphics[width=1.0\linewidth]{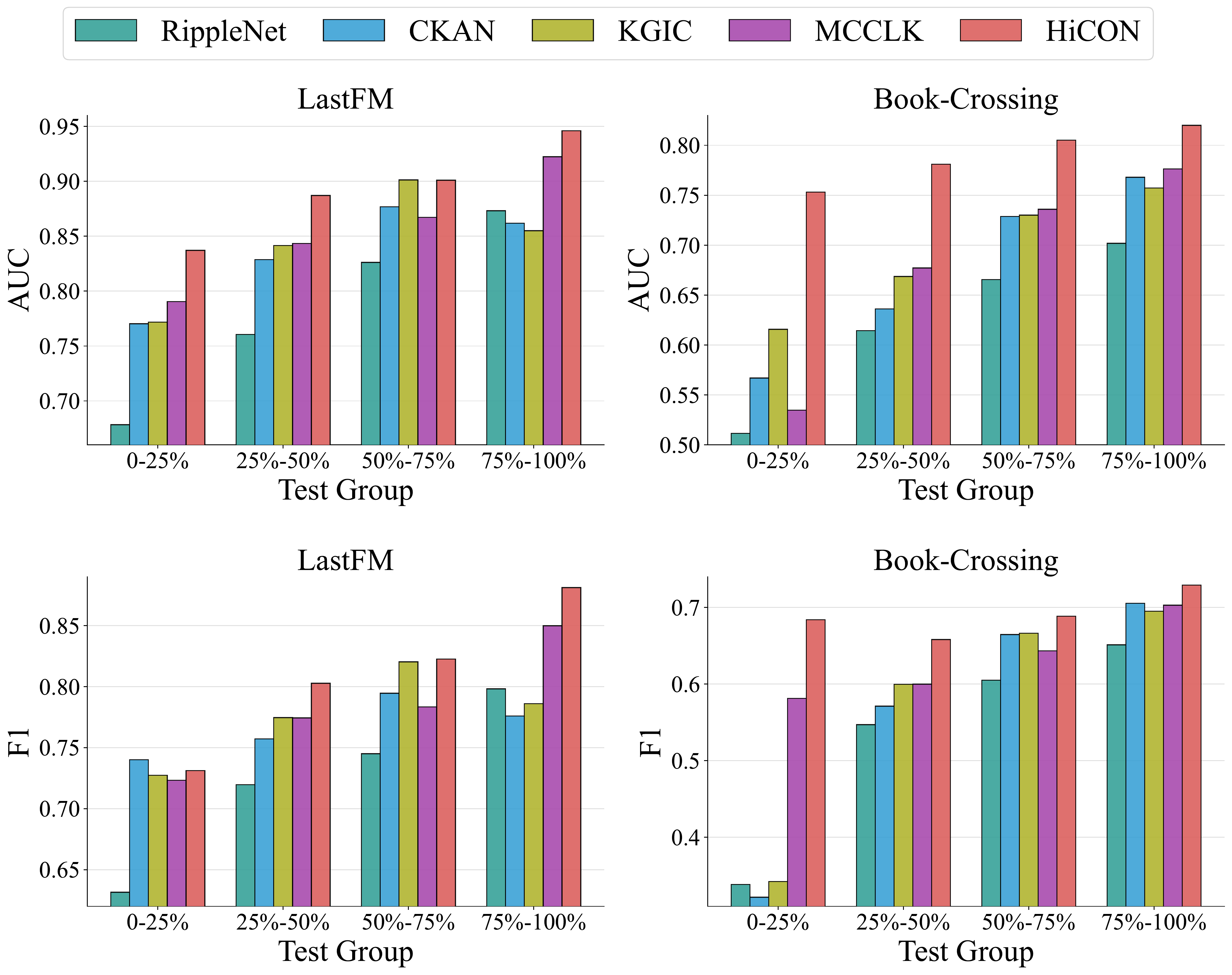}
  \caption{The performance of HiCON and baselines at different interaction sparsity levels. 25\%-50\% is a user group sampled from the test set, including the users whose number of interacted items ranges from the 25th to 50th percentile.}
  \label{Fig:data_sparsity_level}
\end{figure}
\subsection{Performance w.r.t Interaction Sparsity Levels}
In this section, to evaluate the recommendation performance of HiCON at different interaction sparsity levels, we compare HiCON with several representative knowledge-aware baselines, i.e., RippleNet, CKAN, KGIC, and MCCLK.  
Specially, we split the users included in the test set of LastFM into four groups according to the number of their interacted items.
For example, the user group 25\%-50\% contains the users whose interaction number ranges from 25th to 50th percentile. 
The group-wise results are shown in Figure~\ref{Fig:data_sparsity_level}. 
We observe the following findings: 
1) HiCON outperforms most baselines for the inactive users that interact with a small number of items (e.g., group 25\%-50\%).
This may be because our model can effectively exploit more complex semantic relatedness and learn more discriminative representations to alleviate the data sparsity issue;
2) For the active users, HiCON yields consistent improvement over all baselines in terms of AUC and F1. 
This may be because the hierarchical message aggregation selects a bundle of valuable neighbors from vast neighbors and then derives semantic relatedness with them, which is helpful for making accurate recommendations.

\subsection{Hyper-parameter Analysis}
\begin{figure}[t]
  \centering
 \includegraphics[width=1.0\linewidth]{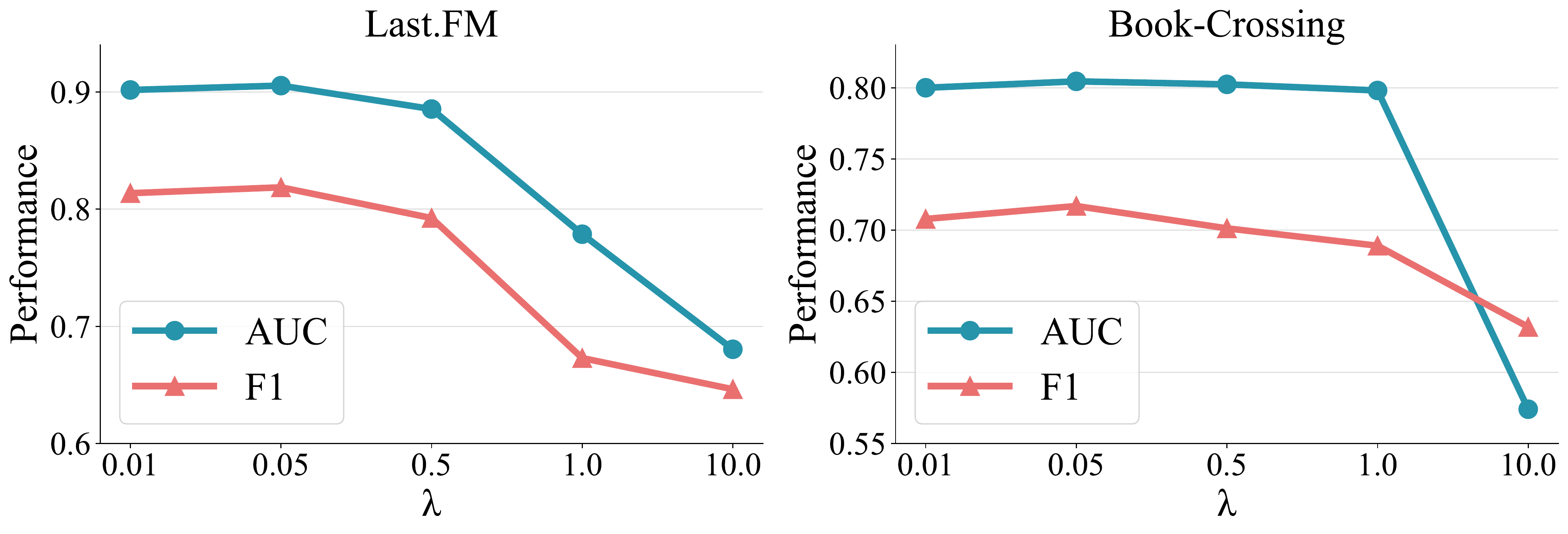}
  \caption{The impact of the hyper-parameter $\lambda$.}
  \label{Fig:sensitivity}
\end{figure}

In this section, we analyze the impact of the hyper-parameter $\lambda$ used for controlling the balance of BPR and contrastive losses. 
The results are shown in Figure~\ref{Fig:sensitivity}. 
We observe that HiCON achieves relatively stable recommendation performance on both AUC and F1 at first, and then its performance declines significantly when $\lambda$ is set to a large number (e.g, 10.0).
This may be because the contrastive learning with a large weight $\lambda$ dominates the overall loss, misleading the model optimization process.

\end{document}